\documentclass[aps,prl,twocolumn,a4paper,10pt,notitlepage,footinbib,superscriptaddress,showpacs]{revtex4-1}%
\usepackage[english]{babel}
\usepackage[T1]{fontenc}
\usepackage{endnotes}
\usepackage{amssymb,amsmath,amsfonts}
\usepackage{textcomp}
\usepackage{graphicx,color}
\usepackage[dvipsnames]{xcolor}
\usepackage[utf8x]{inputenc}
\usepackage{float}

\usepackage{tabularx}
\usepackage{bm}

\newcommand{\dmi}[1]{\textcolor{black}{#1}}%
\newcommand{\dmii}[1]{\textcolor{black}{#1}}%

\renewcommand{\vec}[1]{\boldsymbol{\mathrm{#1}}}%

\begin{document}

\title{Nonequilibrium Theory of Epigenomic \\ Microphase Separation in the Cell Nucleus} 

\vspace{-0.4 cm}
\author{Davide Michieletto}
\email{davide.michieletto@ed.ac.uk}
\thanks{Joint first author}
\affiliation{SUPA, School of Physics and Astronomy, University of Edinburgh, Edinburgh EH9 3FD, United Kingdom}
\affiliation{MRC Human Genetics Unit, Institute of Genetics and Molecular Medicine, University of Edinburgh, Edinburgh EH4 2XU, UK}
\affiliation{Centre for Mathematical Biology, and Department of Mathematical Sciences, University of Bath, North Rd, Bath BA2 7AY, UK}
\author{Davide Col\`i}
\thanks{Joint first author}
\affiliation{Dipartimento di Fisica e Astronomia and Sezione INFN, Università degli Studi di Padova, I-35131 Padova, Italy}
\author{Davide Marenduzzo}
\affiliation{SUPA, School of Physics and Astronomy, University of Edinburgh, Edinburgh EH9 3FD, United Kingdom}
\author{Enzo Orlandini}
\email{orlandini@pd.infn.it}
\affiliation{Dipartimento di Fisica e Astronomia and Sezione INFN, Università degli Studi di Padova, I-35131 Padova, Italy}

\vspace{-0.3 cm}


\begin{abstract}
\vspace{-0.3 cm}
\textbf{Understanding the spatial organisation of the genome in the cell nucleus is one of the current grand challenges in biophysics. Certain biochemical -- or epigenetic -- marks that are deposited along the genome are thought to play an important, yet poorly understood, role in determining genome organisation and cell identity. The physical principles underlying the interplay between epigenetic dynamics and genome folding remain elusive. Here we propose and study a theory that assumes a coupling between epigenetic mark and genome densities, \dmi{and which can be applied} at the scale of the whole nucleus. We show that equilibrium models are not compatible with experiments and a qualitative agreement is recovered by accounting for non-equilibrium processes which can stabilise \dmi{microphase separated epigenomic domains}. We finally discuss the potential biophysical origin of these terms.
}
\vspace{-1 cm}
\end{abstract}

\maketitle

\vspace{-3 cm}
Establishing distinct and inheritable cellular identities in different tissues is crucial to the existence of complex multi-cellular organisms. Because all cells contain the same DNA, cellular fate cannot be directed by genetic cues alone. Instead, tissue-specific 3D genome organisation~\cite{Cavalli2013} and biochemical (also called epigenetic) patterns~\cite{Cortini2015} are thought to be key regulators of cellular fate~\cite{Stadhouders2019}. 
Epigenetic patterns are composed of biochemical marks on DNA and histones -- the proteins which package DNA into chromatin~\cite{Alberts2014} -- and can be deposited or removed by a range of specialised proteins~\cite{Cortini2015} often recruited by complexes such as RNA polymerase~\cite{Ng2003} and Polycomb Repressive Complex~(PRC)~\cite{Zentner2013}. 

Investigating the interplay between genome organisation and epigenetic patterns can shed light into the mechanism underlying cell fate decision~\cite{Pombo2015,Stadhouders2018}. To study the dynamics of epigenetic patterns, Ising-like models have been proposed in the biophysics literature~\cite{Dodd2007,Jost2014pre,Teif2015,Berry2017,Erdel2016,Erdel2018a}. Yet, the genome is poorly represented by a 1D array of Ising spins. Instead, it is a fluctuating polymer which can assume distinct spatial organisations in 3D and also in response to stimuli~\cite{Zirkel2017,Buckle2018} and epigenetic cues~\cite{Stadhouders2019}. 

In this Letter, and the companion paper~\cite{PRE}, we aim to analyse the link between chromatin \dmi{large-scale organisation} and epigenetic dynamics by using models for chromatin folding inspired by the physics of magnetic polymers~\cite{Garel1999a}.  We study a Landau-Ginzburg field theory where the dynamics of epigenetic marks is linked to that of genome folding within the nucleus. Our theory considers a 1D chain of Potts-like spins which is allowed to fluctuate in 3D. It markedly departs from previous works~\cite{Michieletto2016prx,Michieletto2018nar} as we formalise and study a theory for the epigenomic organisation of a whole eukaryotic nucleus and investigate how non-equilibrium processes affect its thermodynamics and kinetics.

We show that while an equilibrium theory for epigenomic organisation captures some features seen {\it in vivo}, such as segregation of different epigenetic marks, it fails to explain the experimentally observed coexistence of diverse epigenetic and genomic domains in eukaryotic nuclei~\cite{Cremer2015,Cortini2015}. In light of this we then propose a non-equilibrium field theory accounting for generic energy-consuming biochemical and biophysical processes in the nucleus. We discover that a simple first-order reaction leads to arrested (and tunable) phase separation of epigenomic domains, in qualitative agreement with experiments~\cite{Cremer2015}. Alongside our theory, we show results from large-scale 3D Brownian Dynamics (BD) simulations of (non-equilibrium) magnetic polymer melts. Besides validating our field theory, they also represent a key step towards a more realistic non-equilibrium polymer model for genome organisation with dynamic epigenetic marks. 
   
\paragraph{Model -- }
Epigenetic marks are biochemical (e.g., methyl) groups, that are transiently deposited on histones or DNA. Albeit a slew of modifications exists~\cite{Alberts2014}, it is typical to consider a generic situation in which two classes of epigenetic states may be present on a given chromatin segment, marking either transcriptionally active or silent DNA~\cite{Cortini2015}\dmi{\cite{Lieberman-Aiden2009}}.

Specific protein complexes can bind and bridge genomic segments bearing the same marks~\cite{Barbieri2012,Brackley2013pnas,Jost2014B} while others are known to deposit biochemical groups~\cite{Hathaway2012}. Importantly, evidence suggest that \dmi{some of the} complexes depositing a given epigenetic mark also contain protein domains which can bind the same mark~\cite{Dodd2007,Cortini2015}~\dmi{(an example is PRC~\cite{Moazed2011,Michieletto2016prx})}.
This read-write mechanism sets up a positive feedback loop which is here captured by a ``ferromagnetic'' interaction tending to align 3D proximal Potts spins -- here playing the role of epigenetic marks -- or to bring them together when already aligned~\cite{PRE}.

We describe the system with two fields $n(\vec{x},t)$ and $m(\vec{x}, t)$ -- the density and ``magnetisation'' fields. The former records the local genomic density at a given spatio-temporal location, the latter \dmi{the local abundance of a given} epigenetic state. These fields should be interpreted as local averages over a volume element centred in $\vec{x}$, large enough to smooth out microscopic fluctuations. Since the total DNA mass is constant during most of the cell cycle, we assume that the mean density $n_0 = \int d\vec{x} \, n(\vec{x},t)/V$ is conserved. A generic Landau-Ginzburg free energy density describing this model is  
\begin{equation}
\beta f_{nuc} =  a m^2 + b m^4 + c n^2 + d n^3 -\chi m^2 n \, .
\label{eq:FMelt}
\end{equation}
The rationale behind Eq.~\eqref{eq:FMelt} is that it must respect the $\mathbb{Z}_2$ symmetry of the magnetisation field (active/inactive chromatin~\cite{Lieberman-Aiden2009}) and it must describe the behaviour of the density field via a standard virial expansion for non-ideal gases~\cite{ChaikinLubensky} so that, for $m=0$, the ground state is a nucleus homogeneously filled with DNA. Hereafter we set $a=b=c=d=1$ unless otherwise stated, thereby suppressing phase transitions driven by any of the two uncoupled fields. The key coupling term, $\chi m^2 n$, is chosen to model the feedback between chromatin folding ($n>0$) and epigenetic ordering ($m^2 > 0$). In this respect, $\chi$ can be thought of as parametrising the self-attraction of equal epigenetic marks, which is mediated by bridging proteins~\cite{Barbieri2012,Brackley2013pnas} such as HP1~\cite{Larson2017,Hiragami-Hamada2016}, polymerases~\cite{Cook2002} and transcription factors~\cite{Sabari2018}.  \dmi{Models implementing this coupling in single chromosomes have been developed in the literature~\cite{Michieletto2016prx,Michieletto2018nar,Michieletto2017scirep,Haddad2017, Jost2018} but no extension to the full nuclear scale has been proposed yet.}
\dmi{For simplicity, Eq.~\eqref{eq:FMelt} neglects the polymeric nature of chromosomes, which is instead clearly captured in the BD simulations (see also SI and Ref.~\cite{PRE}), and also interactions with the nuclear envelope (lamina)~\cite{Solovei2009,Falk2018,Chiang2018}.}

\begin{figure}[t!]
	\vspace{-0.5 cm}
	\centering
	\includegraphics[width=0.44\textwidth]{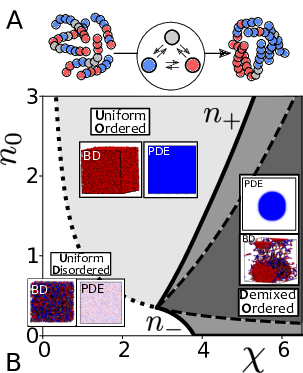}
	\vspace{-0.3 cm}
	\caption{\textbf{Phase Diagram at the Nuclear Scale.} \textbf{A} Sketch of the epigenetic and folding dynamics of magnetic polymers. \textbf{B} Equilibrium phase diagram of the free energy density (Eq.~\eqref{eq:FMelt}). The three equilibrium phases are: (UD) uniform ($n=n_0$) and epigenetically disordered ($m^2=0$); (UO) uniform ($n=n_0$) and epigenetically ordered ($m^2>0$) and (DO) demixed and epigenetically ordered ($n=n_+$, $m^2>0$ and  $n=n_-$,$m^2=0$). 
	The dotted line marks the critical value of the coupling $\chi_c(n_0)=a/n_0$, the solid lines identify the boundaries of the coexistence region (binodals) and the dashed lines identify the spinodal region. Insets report representative snapshots from BD simulations performed at monomer density $\rho$ and self-affinity $\epsilon$ (see SI and Ref.~\cite{PRE} for details). The value of the parameters used for the snapshots are: UD = ($\rho \sigma^3=0.5$, $\epsilon/k_BT=0.7$); UO = ($\rho \sigma^3=0.8$, $\epsilon/k_BT=1$); DO = ($\rho \sigma^3=0.1$, $\epsilon/k_BT=0.9$).  
	See also Supp. Movies \textbf{M1-M2} and \textbf{BD M1-M3}.
	} 
	\vspace{-0.6 cm}
	\label{fig:pd_melt}
\end{figure}

\paragraph{Equilibrium Thermodynamics -- }
We seek to thermodynamically characterise the theory in Eq.~\eqref{eq:FMelt} first by finding an $m$ minimising the free energy for a given $n$, and then by using the common tangent construction to assess whether the resulting uniform state is unstable to phase separation~\cite{Matsuyama2002,Fosado2017}.
The optimal $m^*$ 
is  $m^* = \pm\sqrt{ (\chi n(\vec{x},t) - a)/ 2b}$ if $n(\vec{x},t) > a/\chi$ and 0 otherwise. This solution indicates a second-order phase transition between an epigenetically disordered ($m^2=0$) and ordered ($m^2>0$) phase at the critical line which is shown as a dotted line in Fig.~\ref{fig:pd_melt}. 

Plugging $m^*$ into Eq.~\eqref{eq:FMelt} we obtain the minimised $f^*=f(m^*,n)$, which we need to further minimise with respect to $n$, subject to the constraint that $ \int_V n(\vec{x}) \,\mathrm{d} \vec x/V = n_0$. This can be done
via the common tangent construction~\cite{Fosado2017,Matsuyama2002,ChaikinLubensky}, by finding points in phase space where pressure $P =f^\star - n \partial f^\star/\partial n$ and chemical potential $\mu=\partial f^\star/\partial n$ of the two phases are equal, while having a lower free energy than the mixed/uniform phase. Graphically, these conditions can be solved by finding the points $(n_-, f^*(n_-))$ and $(n_+, f^*(n_+))$ at which the tangents to $f^\star$ have the same slope $\mu$ and the same intercept $P$~\cite{nota1}. By repeating this procedure for different $\chi$ one finds the so-called ``binodals'', $n_-(\chi)$ and $n_+(\chi)$, which are plotted as thick lines in Fig.~\ref{fig:pd_melt}.
Binodal lines delimit the region of phase separation but contain regions in which the mixed state is metastable. The ``spinodal'' region of linear instability, where the system spontaneously demixes into low ($n_-$) and high density ($n_+$) phases, is that where $\partial^2f^\star/\partial n^2 < 0$ yielding $a/\chi < n_0 < (\chi^2-4 b c)/(12 b d)$ shown in Fig.~\ref{fig:pd_melt} as dashed lines.

This construction allows us to discover three possible equilibrium phases for our theory: (i) for $\chi < \chi_c(n_0) = a/n_0$ the system is in a uniform ($n=n_0$) and disordered ($m=0$) phase (UD); (ii) at $\chi=a/n$ we find a second order phase transition to a uniform state ($n(\vec{x},t)=n_0$) with ordered epigenetic field $(m^2>0)$ (UO); (iii) for $\chi > \chi_c (n_+)$ or $\chi > \chi_c (n_-)$, we observe a phase separated state that we call demixed ordered (DO). 

In a biological context, our theory suggests that certain perturbations can trigger nuclear macroscopic phase separation driven by epigenetics. This is achieved either by moving the system into the spinodal region or by overcoming the metastability of the UO phase outside this region. \dmi{Whilst the compactification of large genomic regions (e.g., the inactive X chromosome~\cite{Pinter2012,PRE}) is in line with this result, the uncontrolled spreading of a single epigenetic mark to the whole nucleus is never observed biologically. A possible reason, which we explore in more detail later on, is that the epigenetic read-write dynamics is inherently out of equilibrium.}

\begin{figure}[t!]
\vspace{-0.3 cm}
	\centering
	\includegraphics[width=0.42\textwidth]{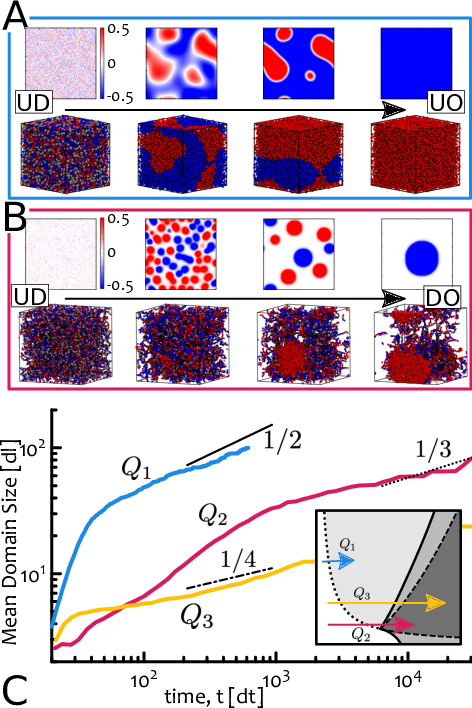}
\vspace{-0.3 cm}
	\caption{\textbf{Dynamics and Reorganisation of Epigenomic Domains.} \textbf{A} Snapshots following the quench $Q_1$ from Uniform Disordered (UD) to Uniform Ordered (UO) ($\chi=0 \rightarrow \chi=1$ at $n_0=2$). \textbf{B} Snapshots following the quench $Q_2$ from Uniform Disordered (UD) to Demixed Ordered (DO) ($\chi=0 \rightarrow \chi=3.5$ at $n_0=0.5$). Top rows in \textbf{A-B} show the evolution of Eq.~\eqref{eq:modelC} on a $100\times 100$ grid. Bottom rows show the evolution through BD simulations (UD$\to$UO, $\rho \sigma^3=0.8$, $\epsilon/k_BT=1$; UD$\to$DO, $\rho \sigma^3=0.1$, $\epsilon/k_BT=0.9$). Ref.~\cite{PRE} gives more details on the BD simulations. \textbf{C} Scaling of the mean epigenetic domain (units of lattice size $dl$) computed as the average size of clusters of connected lattice sites with magnetisation $|m|>1$ for $Q_{1,2}$, and $|m|>2$ for $Q_3$ ($\chi=0 \rightarrow \chi=6$ at $n_0=1$). 
	See also Supp. Movies \textbf{M1-M2} and \textbf{BD M1-M3}. \dmi{The choice of the threshold for $|m|$ is purely for aesthetics and does not change the scaling.} The other parameters are set to $\kappa_m=\kappa_n=5$ and $\Gamma_m=\Gamma_m=0.1$. 
	} 
	\vspace{-0.6 cm}
	\label{fig:dynamics}
\end{figure}
\paragraph{Equilibrium Kinetics -- } 
\dmi{The kinetics of epigenomic reorganisation may be tracked in live cells~\cite{Falk2018} and it could be used as a generic indicator of cell health or to infer the functional implications of intra-nuclear phase separation}. It is therefore important to characterise such kinetics within our theory. To this end we now consider the time-Dependent Landau-Ginzburg equations derived from Eq.~\eqref{eq:FMelt}. These are obtained considering the steepest descent to the free energy minimum for the magnetisation field, $\partial_t m \sim -\delta \mathcal{H}/\delta m$, and a diffusive dynamics of the conserved density, $\partial_t n \sim \nabla^2 \delta \mathcal{H}/\delta n$, with $\mathcal{H} = \int d\vec{x} \left[ f + \kappa_n (\nabla n)^2 + \kappa_m (\nabla m)^2 \right]$~\cite{ChaikinLubensky}. These ``Model C''~\cite{ChaikinLubensky} equations can be explicitly written as
\begin{align}
&  \dot{m} = \Gamma_m \left(2 \chi m n  -2 a m - 4 b m^3 + \kappa_m \nabla^2 m \right)  \notag \\
&  \dot{n} = \Gamma_n \nabla^2 \left( 2 c n + 3 d n^2  - \chi m^2 - \kappa_n \nabla^2 n\right) \, . 
\label{eq:modelC}
\end{align}
We numerically evolve these coupled equations on a $100 \times 100$ grid initialising the system in a UD state with some small fluctuations in both magnetisation and density. From here, a quench $Q$:$\chi \rightarrow \chi^\prime$ is performed and the evolution towards a new stable state monitored.

\begin{figure*}[!t]
	\includegraphics[width=0.85\textwidth]{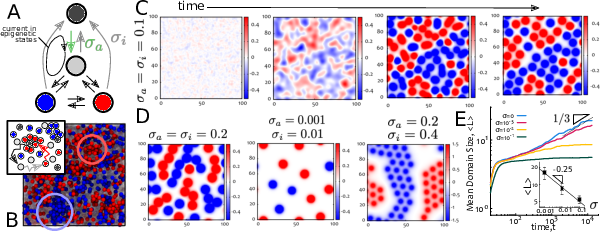}
	\vspace{-0.5 cm}
	\caption{\textbf{Nonequilibrium Chromatin Switching Creates Microphase Separated Epigenomic Domains.} \textbf{A} The switching rates are set so that any part of the genome can become inactive (at rate $\sigma_i$), but an inactive region must reactivate when neutral (at rate $\sigma_a$). This choice sets up a current in epigenetic states that breaks detailed balance and time-reversal symmetry. \textbf{B} Snapshot from a BD simulation of a melt of polymers with switching rate $\kappa_s=10^{-4} \tau_B^{-1}$ ($\tau_B$ is the diffusion time of a monomer~\cite{PRE}), monomer density $\rho=0.8\sigma^{-3}$ and $\varepsilon/k_BT_L=0.9$. The arrows denote a possible loop arising in steady state in BD simulations. \textbf{C} Time-dependent snapshots obtained evolving Eqs.~\eqref{eq:modelC+switch} with $\Gamma_m=\Gamma_n=0.1$, $k_m=k_n=5$, $\chi = 6$, $n_0=1$, $\sigma_{a} =\sigma_i = 0.1$. \textbf{D} Steady state configurations obtained with same parameters except for the switching rates, which are given in the figure.  \textbf{E} Log-log plot of mean epigenetic domain size as a function of time and for different switching rates. The inset shows (again in log-log) the dependence of the average size in steady state against $\sigma=\sigma_a=\sigma_i$. See also movies \textbf{M3-M9}.
	}
	\vspace*{-0.5 cm}
	\label{fig:Switching}
\end{figure*}

Interestingly, the quench into the UO state ($Q_1$ in Fig.~\ref{fig:dynamics}) follows a different scaling with respect to all the others. In $Q_1$ the density field remains uniform while the epigenetic domains form bicontinuous spanning clusters evolving into one percolating domain (see {\bf Movie M1}). The scaling of the typical domain size grows as $L(t) \sim t^{\alpha}$ where $\alpha = 0.46$ is compatible with the coarsening of a non-conserved field such as the magnetisation in the Ising model~\cite{Kockelkoren2002}. 
In all other quenches there is reorganisation of both epigenetic and density fields and our theory yields a slower growth with $\alpha \simeq 0.25 - 0.3$. The lower end is interpreted as a transient behaviour which would eventually lead to the $1/3$ exponent expected in density-conserving Model B kinetics~\cite{Kockelkoren2002} (see \textbf{Movie M2}). It would be interesting to follow the growth of epigenomic domains in live cells, e.g. in oncogene-triggered senescence~\cite{Boumendil2019,Chiang2018} \dmi{or during nuclear inversion~\cite{Falk2018} to compare against the predictions of this model. }

\dmi{Clustering and condensates of proteins, which may be due to liquid-liquid or polymer-polymer phase separation~\cite{Erdel2018a}, and displaying a coherent epigenetic mark on the locally recruited chromatin~\cite{Cortini2015,Moazed2011,Zentner2013} are commonly observed in the nucleus (e.g. Polycomb bodies~\cite{Wani2016}, transcription factories~\cite{Cook2002} and transcriptional condensates~\cite{Cho2018,Sabari2018})}. Their coarsening is arrested and does not proceed indefinitely as predicted by Eqs.~\eqref{eq:modelC}; for this reason, we now aim to improve our model and account for non-equilibrium processes.

\paragraph{Non-equilibrium Model -- } 
\dmi{Models implementing non-equilibrium reactions have been developed for, e.g., the formation and maintenance of centrosomes~\cite{Zwicker2014, Zwicker2018}, stress granules~\cite{Wurtz2018}, nucleoli~\cite{Berry2015a} and the organisation of chromatin~\cite{Hilbert2018} and chromosomes~\cite{Fudenberg2016, Gibcus2018, Brackley2017biophysj, Brackley2017prl}, \dmii{and some of these predict the arrest of large-scale phase separation of nuclear and cytosol components~\cite{Hilbert2018,Zwicker2014}}. Typical processes involved are post-translational modification of nucleosomes and proteins~\cite{Wang2013,Li2012a}, ATP-dependent chromatin remodelling~\cite{Brangwynne2015}, disaggregases~\cite{Narayanan2019} and transcription~\cite{Henikoff2008,Berry2015a,Hilbert2018}}. Some of them can affect the local chromatin state and its accessibility to proteins~\cite{Brangwynne2015} and under certain conditions, they may destabilise or hinder the deposition of histones with a given epigenetic mark~\cite{Henikoff2008}.

\dmi{We propose to implement these effects using a generic ``epigenetic switch'' that can dynamically convert the density field $n$ between a ``magnetisable'' chromatin state that can be epigenetically marked ($n_a$), and an ``inert'' state that cannot be modified ($n_i$). This switch may be viewed as a generic and simplified way to account for ATP-dependent remodelling: inert states are segments of the genome that are refractive to epigenetic modifications (because inaccessible or poor in nucleosomes) while magnetisable ones can receive epigenetic marks. This simple first-order reaction (Fig.~\ref{fig:Switching}A) can also effectively describe a situation in which a genomic site is dynamically associated with a protein that can either strongly ($n_a$) or weakly ($n_i$) bind another genomic site with the same mark.} 
The dynamical equations with epigenetic switching are
\begin{align}
&  \dot{m} = \Gamma_m \left(2 \chi m n_a -2 a m - 4 b m^3 + \kappa_m \nabla^2 m \right)  \notag \\
&  \dot{n}_a = \Gamma_n \nabla^2 \left(g(n_a) - \chi m^2 - \kappa_n \nabla^2 n_a\right)  + \sigma_{a} n_i - \sigma_{i} n_a  \notag \\
& \dot{n}_i = \Gamma_n \nabla^2 \left(g(n_i) -\kappa_n \nabla^2 n_i\right) -\sigma_{a} n_i + \sigma_{i} n_a
\label{eq:modelC+switch}
\end{align}
where $g(x)=2 c x + 3 d x^2$ and the parameters $\sigma_{a/i}$ describe the rates at which chromatin is activated/inactivated (in general $\sigma_i \neq \sigma_a$). Note that we now require the {\it sum} of the two density fields to be conserved~\cite{notaglotzer}. We numerically evolve Eqs.~\eqref{eq:modelC+switch} on a $100 \times 100$ grid starting from the UD phase (other parameters are given in Fig.~\ref{fig:Switching}).   

The rates are set so that any region of the system with $m\ne 0$ can become \dmi{inert}; yet, because the field $n_i$ decouples from $m$, \dmi{the return to a magnetisable state} occurs through a neutral ($m=0$) state. This sets up a current in epigenetic states which breaks detailed balance and time-reversal symmetry (Figs.~\ref{fig:Switching}A,B), similarly to what occurs in theories for scalar active matter~\cite{ElsenPRX}. The current is \dmi{more clearly} visible in our BD simulations, when following the evolution of a chromatin bead with a given epigenetic mark (Fig.~\ref{fig:Switching}B, inset). 

The non-equilibrium switching terms in Eqs.~\eqref{eq:modelC+switch} yield an arrested phase separation with multiple epigenomic domains in steady state, or microphase separation, as in the majority of cell types~\cite{Cremer2015}. We note there is comparatively much less density variations with respect to equilibrium (Figs.~\ref{fig:Switching}B,C).
For large $\sigma_{a/i}$, chromatin droplets of the same magnetisation form ``super-beads'' which come together forming branched aggregates (Fig.~\ref{fig:Switching}D), as this configuration minimises interfaces in the fields $m$ and $n_{i}$ (see Movies \textbf{M6}-\textbf{M9}). The domain size in steady state is controlled by the switching rate, and we find it approximately decays as $\sigma^{-1/4}$ (Fig.~\ref{fig:Switching}E). 

Large-scale BD simulations of magnetic polymer melts in which beads switch between a inert and magnetisable states  at rate $\kappa_s$ confirm that chromatin switching can yield microphase separation of epigenomic domains (Fig.~\ref{fig:Switching}B and Ref.~\cite{PRE}). Besides switching, other ingredients may be at play, such as effective emulsification by RNA~\cite{Hilbert2018}, RNA-binding gel-forming proteins~\cite{Nozawa2017,Michieletto2019rnareview} epigenetic bookmarking~\cite{Michieletto2018nar}, and differential underlying transcriptional activity~\cite{Smrek2017}. We note that copolymers made by heterochromatin and euchromatin would also generically microphase separate~\cite{Leibler1980b,Jost2014B,Brackley2013pnas,Scolari2015} in equilibrium -- yet the epigenetic marks and the emerging domains would be static, unlike those found here.
 \dmi{Finally, we note that a range of mechanisms can lead to the arrest of domain coarsening in generic field theories~\cite{Glotzer1995,Aggeli2001,Weber2019a} and pinpointing the biophysical principles at play in the cell nucleus is a current major challenge}.

\paragraph{Conclusions -- } 
In summary, we have proposed a Landau-Ginzburg theory to describe the coupling between the dynamics of epigenetic states and that of the genome within the cell nucleus. By combining dynamical mean field theory and BD simulations, we find that this theory leads to a growth of competing epigenetic domains, in agreement with the well-known spreading of certain histone modifications~\cite{Hathaway2012}. Our theory predicts several dynamical exponents that can be tested experimentally by triggering epigenomic reorganisations. 

In equilibrium, one dominant epigenetic domain eventually takes over the whole nucleus. \dmi{This is in stark contrast with experiments, where multiple stable domains coexist}. We found that a generic nonequilibrium ``epigenetic switch'' between modifiable and refractory chromatin states (or first-order biochemical reaction) can arrest the unlimited spreading of epigenetic marks and stabilise a microphase separated state with coexisting epigenomic domains. 
\dmi{It would be of interest to complement or refine our models with more specific non-equilibrium processes such as RNA production~\cite{Hilbert2018,Berry2015a}, phosphorylation or ATP-controlled switching of protein conformations or interactions~\cite{Brangwynne2015,Brackley2017biophysj,Nozawa2017,Michieletto2019rnareview}.} We hope that our predictions will be tested in vivo and in vitro by controlling the rates of post-translational modification of certain protein complexes that are known to bind and/or remodel chromatin.

\paragraph{Acknowledgements -- } We thank the European Research Council (ERC CoG 648050 THREEDCELLPHYSICS) for funding.   DMi and EO would also like to acknowledge the networking support by EUTOPIA (CA17139).

\bibliographystyle{apsrev4-1}
\bibliography{library,nota1,nota2,notaglotzer}

\begin{thebibliography}{71}
\bibitem [{not({\natexlab{a}})}]{nota1}%
  \BibitemOpen
  \href@noop {} {} ({\natexlab{a}}),\ \bibinfo {note} {we note that for our
  chosen parameters and $\chi \geq 3.8$ this construction would lead to
  $n_-<0$, which is non-physical. For this reason one needs to impose that
  $n_-=0$ and that the slope (chemical potential $\mu_-$) at $n_-$ can attain
  any value. By doing so we end up with one equation for $n_+$, which is the
  one imposing the condition that the pressure (intercept) is null, i.e.
  $f^*-n_+\mu_+=0$.}\BibitemShut {Stop}%
\bibitem [{not({\natexlab{b}})}]{notaglotzer}%
  \BibitemOpen
  \href@noop {} {} ({\natexlab{b}}),\ \bibinfo {note} {if uncoupled from the
  magnetic field, and $\sigma_i=\sigma_a$, this model could be re-written as an
  effectively equilibrium theory for the density field with short and long
  range interactions~\cite{Glotzer1994,Glotzer1995}.}\BibitemShut {Stop}%
\end{thebibliography}

\end{document}